\documentclass[reprint,amsmath,amssymb,aps,eqsecnum,showpacs,prx]{revtex4-1}
\usepackage{graphicx}
\usepackage{bm}
\usepackage{epsfig}
\usepackage{amssymb}
\usepackage{amsfonts}
\usepackage{braket}
\usepackage{color}
\usepackage{epstopdf}
\usepackage{hyperref}
\usepackage{float}
\restylefloat{table}
\usepackage{bibentry}
\usepackage{multirow}
\usepackage[caption=false]{subfig}
\usepackage{braket}
\newcommand{\ba}{\begin{eqnarray}}
\newcommand{\ea}{\end{eqnarray}}
\newcommand{\bd}{\begin{displaymath}}
\renewcommand{\v}[1]{{\bf #1}}
\newcommand{\nn}{\nonumber \\}
\DeclareGraphicsExtensions{.pdf,.png,.jpg}
\graphicspath{{C:\Users\Home\Dropbox\Lab\AKLT}}

\begin{document}
\title{Low-energy dynamics of the Affleck-Kennedy-Lieb-Tasaki model \\ in the one- and two-triplon basis}

\author{Jintae Kim}
\affiliation{Department of Physics, Sungkyunkwan University, Suwon 16419, Korea}
\author{Rajarshi Pal}
\affiliation{Department of Physics, Sungkyunkwan University, Suwon 16419, Korea}
\author{Jung Hoon Han}
\email[Electronic address:$~~$]{hanjemme@gmail.com}
\affiliation{Department of Physics, Sungkyunkwan University, Suwon 16419, Korea}
\date{\today}
\begin{abstract} The elementary excitation in the antiferromagnetic spin-1 model known as the Affleck-Kennedy-Lieb-Tasaki (AKLT) Hamiltonian has been described alternatively as magnons or kink-like solitons (triplons). The latter, which we call the triplon throughout this paper, has been proven equivalent descriptions of the same magnon excitation and not an independent branch of excited states. On the other hand, no careful examination of multi-magnon and multi-triplon equivalence was made in the past. In this paper we prove that two-magnon and two-triplon states are also identical descriptions of the same excited states, and furthermore that their energies break down as the sum of one-triplon energies {\it exactly} for the AKLT Hamiltonian. The statement holds despite the fact that the model is non-integrable. Such magnon/triplon dichotomy is conjectured to hold for arbitrary $n$-magnon and $n$-triplon states. The one- and two-triplon states form orthogonal sets that can be used to span the low-energy Hilbert space. We construct an effective version of the AKLT Hamiltonian within such subspace, and work out the correction to the one-triplon energy gap that finds excellent agreement with the known exact value.
\end{abstract}
\maketitle

\section{Introduction}

A one-dimensional chain of quantum-mechanical spins of size $S$ has proven a rich source of insights regarding the nature of low-dimensional quantum spin states and their (sometimes) exotic excitations. The method by which to solve the $S=1/2$ antiferromagnetic chain problem was laid out by Bethe~\cite{bethe31} in the early days of quantum mechanics, but it was not until 1981 that the exotic elementary excitations of the ground state known as spinons - fermions with spin quantum number 1/2 - were discovered by Faddeev and Takhtajan~\cite{faddeev81}. At around the same time, Haldane~\cite{haldane83,haldane83-1} and soon afterwards Affleck, Kennedy, Lieb, and Tasaki (AKLT)~\cite{AKLT87,AKLT88} pioneered the physics of $S=1$ antiferromagnetic spin chain and its crucial differences with the $S=1/2$ chain.

In a variational approach, Arovas, Auerbach, and Haldane proposed a spin wave excitation carrying the spin quantum number $\Delta S=1$ and computed its energy within the single-mode approximation (SMA) scheme~\cite{arovas88}. The one-magnon spectrum thus obtained matched the excitation spectrum calculated from the exact diagonalization of the $S=1$ spin chain model around the momentum $k=\pi$~\cite{solyom93}. Arovas~\cite{arovas89} subsequently came up with ways to construct ``exact excited states" of the AKLT Hamiltonian with the total spin $\Delta S=1$, and 0, respectively, and Regnault {\it et al.}~\cite{bernevig18a} made exact excited states for higher total spins. These are quite high-energy excitations, however, that presumably have little relevance to low-energy dynamics of the integer spin chain. Soon it began to emerge that a string-like triplon construction of the excited state was also possible, but that it gave rise to the identical excitation spectrum within SMA as the magnon construction because their respective wave-like superpositions actually resulted in the same wave function~\cite{knabe88,solyom93,totsuka95,mikeska95}. In recent years, sophisticated numerical implementation of the matrix product state (MPS) algorithm has shed light on the physics of spin-1 chain~\cite{verstraete13,verstraete15,verstaete16,Klumper93,Bartel03}. In the mean time, the challenging problem of constructing a two-magnon or two-triplon wave function and calculating their energies has never been taken up. Here we address these problems explicitly for the AKLT Hamiltonian, motivated by the classic work by Dyson on the multi-magnon construction for the ferromagnet~\cite{dyson56}. One of the innovations made in Dyson's paper is the construction of mutually orthogonal multi-magnon basis states that are also nearly free, in the sense that the scattering matrix elements between magnons become vanishingly small in the long-wavelength limit.

Following the spirit of Dyson's analysis, we construct two-magnon and two-triplon excited states atop the AKLT ground state, and work out their properties. Firstly we show that the two constructions of the excited states are identical, and extend the previously known equivalence of one-magnon and one-triplon excited states~\cite{solyom93,totsuka95} to the two-magnon case. Employing the SMA scheme, we compute the two-triplon energy using the AKLT Hamiltonian, to find that its energy is the sum of the one-triplon energies without further correction from interaction effects. Finally we prove that one- and two-triplon states are orthogonal and can be used to construct a low-energy subspace of the AKLT Hamiltonian. The effective Hamiltonian of the AKLT model within such subspace is constructed, for future use in the investigation of dynamics in the AKLT chain. As a preliminary example of such investigation, we compute the one-triplon energy gap using the effective AKLT Hamiltonian and find improved agreement with the exact value.

Due to the long history of literature on the AKLT model and its excited states, we find it convenient to review important past findings in Sec. \ref{sec:review} and develop our own contributions from there on. Construction of two-magnon and two-triplon states and the proof of their equivalence is given in Sec. \ref{sec:two-triplon}. Calculation of their energies is performed as well. Orthogonality of the one- and two-triplon states and the construction of effective Hamiltonian within such subspace are presented in Sec. \ref{sec:effective-H}. Our paper contains extensive amount of computations of matrix elements and other technical details, which can be found in several Appendices.

\section{One-magnon/one-triplon dichotomy in the AKLT model}
\label{sec:review}

Schwinger boson formalism provides great advantage in understanding the ground state and the low-lying excited states of the AKLT model Hamiltonian~\cite{arovas88,arovas89}. The boson substitution of the spin operator takes place as
\begin{align}
S_{i}^{z} &= \frac{1}{2}(a_{i}^{\dagger}a_{i} - b_{i}^{\dagger}b_{i}) & S_{i}^{+} &= a_{i}^{\dagger}b_{i} & S_{i}^{-} &= b_{i}^{\dagger}a_{i}
\end{align}
with a pair of bosons $a_i$ and $b_i$ at every site $i$. For $S=1$, the three allowed spin-1 states are
\begin{align}\label{eq:rep}
\frac{1}{\sqrt{2}} (a_{i}^{\dagger} )^2 \ket{v} &= \ket{1} &
a_{i}^{\dagger}b_{i}^{\dagger}\ket{v} &= \ket{0} &
\frac{1}{\sqrt{2}} ( b_{i}^{\dagger} )^2 \ket{v} &= \ket{-1} \nonumber
\end{align}
where $|v \rangle$ is the vacuum of the Schwinger bosons, and $+1,0,-1$ are the three allowed orientations of spin-1. The AKLT state has an intuitive expression~\cite{arovas88} in the boson language
\begin{equation}
\ket{A} = \left( \prod_{i} {\cal S}^\dag_{i,i+1} \right) \ket{v} .  \label{eq:AKLT-in-SB}
\end{equation}
The operator ${\cal S}^\dag_{i,i+1} = a_{i}^{\dagger}b_{i+1}^{\dagger}-a_{i+1}^{\dagger}b_{i}^{\dagger}$ creates a singlet pair on the neighboring $\langle i,i+1 \rangle$ sites. The AKLT Hamiltonian $H_{A} = \sum_i H_i$ is given as the sum of projectors
\ba
H_i
&=& {\frac{1}{24}} [ \v S_{i} +  \v S_{i+1}  ]^2 ( [ \v S_{i} +  \v S_{i+1}  ]^2  - 2)  \nn
&=& \frac{1}{24}(  [ {\cal S}_{i,i+1}^{\dagger} ]^2{\cal S}_{i,i+1}^{2}  - 6 { \cal S}_{i,i+1}^{\dagger}{\cal S}_{i,i+1}  + 24) . \label{eq:H-AKLT} \ea
The last line follows from the easily verified identity $
\v S_{i}\cdot\v S_{i+1}= -\frac{1}{2}{\cal S}_{i,i+1}^{\dagger}{\cal S}_{i,i+1} + 1$.

Now, rather than having a singlet, the $(ij)$ bond may have one of the triplet states given by
\ba  {\cal T}^{0}_{ij} &  = & a_{i}b_{j}+a_{j}b_{i}  \nn
{\cal T}^{~1}_{ij}  & = & a_{i} a_{j}\nn
{\cal T}^{~ -1}_{ij}  &  = &  b_{i} b_{j} .
\ea
The superscript refers to the quantum numbers of the triplet excitation. In the subquent notation we will use ${\cal T}^{\pm}$ interchangeably with ${\cal T}^{\pm 1}$. The one-triplon state $|{\cal T}_i^{\alpha}\rangle$ is defined as the one in which the singlet bond ${\cal S}^\dag_{i,i+1}$ in the AKLT state is replaced by one of the triplet creation operators $[{ \cal T}^{\alpha}_{i,i+1} ]^\dag$. The one-triplon states are orthogonal to the AKLT state, $\braket{A|{\cal T}_{i}^{\alpha} } = 0$, as can be proven by explicit calculation or through the symmetry argument that the two states carry different quantum numbers. For a pair of one-triplon states when $i>j$ one can prove~\cite{solyom93}:
\ba\label{eq:1TO}
\braket{{\cal T}_i^{\alpha_1 }|{\cal T}_j^{\alpha_2 }}
=\delta_{\alpha_1,\alpha_2}\frac{1+\delta_{\alpha_1,0}}{2}\left((-1)^N(-\frac{1}{3})^{-i+j}+3^N(-\frac{1}{3})^{i-j}\right)\nn
\ea
The method by which this overlap and many other overlaps in this paper have been tabulated is the transfer matrix method, and one can find its details in the Appendix \ref{appendix:TM}. In the above, $N$ stands for the length of the chain.

The one-triplon state, which seems like a local excitation in the bond operator language, is actually a highly non-local excitation in the spin language as the following identity testifies~\cite{knabe88,solyom93,totsuka95}:
\ba |{\cal T}_i^{\alpha} \rangle = \sum_{j \le i} 2|S^{\alpha}_j \rangle  ~~ (\alpha=x,y,z) .  \label{eq:triplon-vs-magnon} \ea
Here $|S^{\alpha}_j \rangle \equiv S_j^{\alpha}|A\rangle$ is a shorthand for local spin excitation at site $j$. The $x,y$ components of the triplet excitation is defined by $|{\cal T}_i^{\pm } \rangle = \mp ( |{\cal T}_i^{x} \rangle \pm  i |{\cal T}_i^{y} \rangle )/2$, hence $|{\cal T}_i^{\pm } \rangle =\mp  \sum_{j \le i}|S^{\pm }_j \rangle$ $(S^\pm = S^x \pm i S^y )$. The $z$-component of the triplet is a re-definition: $| {\cal T}^z \rangle = |{\cal T}^0 \rangle$. Clearly, the triplon is a solitonic operator affecting all spins to the left of its site of creation.

An interesting consequence follows from forming the triplon wave, with momentum $k$,
\begin{equation}
\ket{{\cal T}^{\alpha}_k }=\sum_{i}e^{ikx_{i}}\ket{{\cal T}^{\alpha}_{i}} .
\end{equation}
By a simple calculation one can show
\ba \ket{ S^\alpha_k } &\equiv & \sum_i e^{ikx_i} \ket{ S^\alpha_i }
= \frac{1}{2} (1-e^{ik} ) \ket{ {\cal T}^\alpha_k }  ~~ (\alpha=x,y,z) , \nn
\ket{ S^{\pm}_k }  &=& \mp (1-e^{ik}) \ket{ {\cal T}^\pm_k } .  \label{eq:Tk-vs-Sk} \ea
Hence, the one-magnon wave and the one-triplon wave states are identical~\cite{solyom93}. The energy of either of these states is computed as~\cite{arovas88}
\ba \label{eq:SMA}
\omega_1 (k) &=& \frac{  \braket{{\cal T}^\alpha_k |H_{A}|{\cal T}^\alpha_k  }}{\braket{{\cal T}^\alpha_k  | {\cal T}^\alpha_k } }  = \frac{5}{27} (5 + 3 \cos k) ,\ea
in the SMA. The one-magnon (one-triplon) energy is independent of $\alpha$ due to the spin-rotation invariance of the Hamiltonian. The excitation is triply degenerate. The one-magnon (one-triplon) SMA spectrum describes the exact excited state spectrum near $k=\pi$ with high accuracy~\cite{solyom93}.

We leave this section with a minor comment. At $k=0$, Eq. (\ref{eq:Tk-vs-Sk}) becomes ill-defined and we also have $\ket{S^{\alpha}_{k=0}}=\sum_i \ket{S^\alpha_i} = 0$ because the AKLT state is a spin-singlet. As a result, we may apply the $k\rightarrow 0$ limiting procedure to (\ref{eq:Tk-vs-Sk}) to obtain the $k=0$ triplon state as
\ba
|{\cal T}^\alpha_{k=0}\rangle = \lim_{k \to 0} \frac{\sum_i e^{ikx_i}S^\alpha_i\ket{A} }{e^{ik}-1} = \sum_i x_i \ket{S^\alpha_i }.
\ea
Despite the appearance of position coordinate $x_i$, this is a translationally invariant state due to the periodic boundary condition of the lattice $i+N \equiv i$.

\section{Two-magnon and two-triplon excitations}
\label{sec:two-triplon}

In the previous section we mentioned that the one-magnon spectrum computed in the SMA describes the energies of the exact excited states fairly well around $k=\pi$. It was also known for some time that the $k\approx 0$ spectrum is reasonably well-described as two independent one-magnon excitations at the respective momenta $\pi+k/2$ and $-\pi+k/2$~\cite{arovas89}:
\ba
\omega_{2} (k)  &=& \omega_1\left( \pi+k/2 \right)+\omega_1 \left( -\pi+k/2 \right) \nn
&=& \frac{10}{27}\left( 5-3\cos \frac{k}{2}\right) . \label{eq:omega2int}
\ea
If the magnons were truly independent and devoid of interactions, the multi-magnon energy is naturally given as the sum of the one-magnon energies, but such assumptions must be subject to careful scrutiny, especially given the fact that the AKLT model is non-integrable. We address the question of two-magnon energy by first constructing a two-magnon state, then evaluating its energy in the manner of SMA. Specifically, we show that the two-magnon wave states at momenta $(k_1, k_2)$ have the SMA energy given exactly as the sum of two one-magnon energies, $\omega_2 (k_1 , k_2 ) = \omega_1 (k_1 ) + \omega_1 (k_2 )$, devoid of any further corrections. The formula (\ref{eq:omega2int}) is a special case of this general conclusion. The other accomplishment of this section is the proof that the two-triplon wave state, appropriately constructed, is again identical to the two-magnon wave states at the same momenta $(k_1 , k_2)$.

\subsection{Equivalence of two-magnon and two-triplon wave states}

First of all, let us define the two-magnon and the two-triplon states, respectively. The two-triplon state replaces two of the singlet bonds in the AKLT state by triplets, at the $(i,i+1)$ and $(j,j+1)$ bonds. When the two sites coincide $i=j$ the wave function vanishes automatically as it violates the constraint of the total spin $S=1$ at the site. The two-triplon states are denoted $\ket{ {\cal T}^{\alpha}_i {\cal T}^{\beta}_j }$. The pair of indices $\alpha, \beta = \pm , 0$ refer to the spin quantum number. We will also refer to them as angular momentum quantum numbers from time to time. The two-magnon state, on the other hand, is defined as a pair of spin operators acting on two sites: $ \ket{ S^{\alpha}_{i} S^{\beta}_{j}} = S^{\alpha}_{i} S^{\beta}_{j} |A\rangle$. As is clear from their respective definitions, they are very different kinds of excitations, one being non-local and the other being local in nature. Now we wish to ask if some correspondence exists between their respective wave-like states given as follows:
\ba
\ket{S_{k_1}^{\alpha_1}S_{k_2}^{\alpha_2}}&=&\sum_{i,j}e^{ik_1x_i+ik_2x_j} \ket{ S^{\alpha_1}_iS^{\alpha_2}_j }\nn
\ket{{\cal T}_{k_{1}}^{\alpha_1}{\cal T}_{k_{2}}^{\alpha_2}} &=&\sum_{i,j}e^{ik_1x_i+ik_2x_j}\ket{{\cal T}^{\alpha_1}_i{\cal T}^{\alpha_2}_j} . \label{eq:wave-definitions}
\ea
We dub them as two-magnon waves and two-triplon waves, respectivesly.

By application of identities such as
\ba
S_i^-\ket{{\cal T}_i^{+}} & = & \frac{1}{2}\ket{{\cal T}_i^{0}}-\frac{1}{2}\ket{A}-\ket{{\cal T}_{i-1}^{-}{\cal T}_i^{+}}\nn
S_{i+1}^-\ket{{\cal T}_i^{+}} & = &\frac{1}{2}\ket{{\cal T}_i^{0}}+\frac{1}{2}\ket{A}-\ket{{\cal T}_{i}^{+}{\cal T}_{i+1}^{-}}\nn
S_{i+1}^+\ket{{\cal T}_i^{+}}&=&-\ket{{\cal T}_i^{+}{\cal T}_{i+1}^{+}}\nn
S_{i}^+\ket{{\cal T}_i^{+}}&=&\ket{{\cal T}_{i-1}^{+}{\cal T}_{i}^{+}}
\ea
one can work out the relation between the two-magnon and two-triplon wave states. Introducing a mnemonic $R(k_1, k_2) = (e^{ik_1} - 1)(e^{ik_2} -1)$, we find a set of {\it exact} relations between two-magnon and two-triplon waves
\begin{widetext}
\ba
\ket{S_{k_1}^\pm S_{k_2}^\pm }&=&R (k_1, k_2 ) \ket{{\cal T}_{k_{1}}^\pm  {\cal T}_{k_{2}}^\pm } \nn
\ket{S_{k_1}^z S_{k_2}^z}&=&\frac{1}{4} R(k_1 , k _2 ) \Bigl(\ket{{\cal T}_{k_{1}}^0 {\cal T}_{k_{2}}^0 }+N\delta_{k_1,-k_2}\ket{A}\Bigl)\nn
\ket{S_{k_1}^\pm  S_{k_2}^\mp  }&=&- R(k_1 , k_2 ) \Bigl(\ket{{\cal T}_{k_{1}}^\pm  {\cal T}_{k_{2}}^\mp } \pm  \frac{1}{2}\frac{e^{ik_1}+1}{e^{ik_1} - 1}\ket{{\cal T}_{k_1+k_2}^0 }-\frac{1}{2}N\delta_{k_1,-k_2}\ket{A}\Bigl)\nn
\ket{S_{k_1}^\pm  S_{k_2}^z } &=& \mp  \frac{1}{2} R(k_1 , k_2 ) \Bigl(\ket{{\cal T}_{k_{1}}^\pm   {\cal T}_{k_{2}}^0 } \pm \frac{e^{ik_1}+1}{e^{ik_1}-1}\ket{{\cal T}_{k_1+k_2}^\pm  }\Bigl)\nn
\ket{S_{k_1}^z S_{k_2}^\pm  }&=&\mp  \frac{1}{2} R(k_1 , k_2 ) \Bigl(\ket{{\cal T}_{k_{1}}^0 {\cal T}_{k_{2}}^\pm  } \mp \frac{e^{ik_1}+1}{e^{ik_1}-1}\ket{{\cal T}_{k_1+k_2}^\pm  }\Bigl) . \label{eq:correspondence}
\ea
\end{widetext}
In general the right-hand side consists of two-triplon, one-triplon, and AKLT states. However, one must be careful to note that the states in these equations are un-normalized. In fact, a careful calculation of the overlaps shows $\langle {\cal T}^{\alpha_1}_{k_1} {\cal T}^{\alpha_2}_{k_2} | {\cal T}^{\alpha_1}_{k_1} {\cal T}^{\alpha_2}_{k_2} \rangle \sim O (3^N \cdot N^2 )$ but the single-triplon and AKLT overlaps  are only
\ba
\braket{ {\cal T}_{k}^{\alpha} | {\cal T}_{k}^{\alpha} } = \frac{2(1+\delta_{\alpha,0})}{5+ 3 \cos k} 3^N N \label{eq:one-triplon-overlap}
\ea
and $\langle A | A \rangle = 3^N$, respectively. Due to the factor $N$ multiplying the AKLT wave function in Eq. (\ref{eq:correspondence}), the two-triplon and AKLT components are of comparable weights, but the one-triplon component has vanishing weight in the thermodynamic limit $N\rightarrow\infty$. If we neglect the one-triplon part for such reason, the above formula simplifies to
\ba
\ket{S_{k_1}^\pm S_{k_2}^\pm }&=&R (k_1, k_2 ) \ket{{\cal T}_{k_{1}}^\pm  {\cal T}_{k_{2}}^\pm } \nn
\ket{S_{k_1}^z S_{k_2}^z}&=&\frac{1}{4} R(k_1 , k _2 ) \Bigl(\ket{{\cal T}_{k_{1}}^0 {\cal T}_{k_{2}}^0 }+N\delta_{k_1,-k_2}\ket{A}\Bigl)\nn
\ket{S_{k_1}^\pm  S_{k_2}^\mp  }&=&- R(k_1 , k_2 ) \Bigl(\ket{{\cal T}_{k_{1}}^\pm  {\cal T}_{k_{2}}^\mp } -\frac{1}{2}N\delta_{k_1,-k_2}\ket{A}\Bigl)\nn
\ket{S_{k_1}^\pm  S_{k_2}^z } &=& \mp  \frac{1}{2} R(k_1 , k_2 ) \ket{{\cal T}_{k_{1}}^\pm   {\cal T}_{k_{2}}^0 } \nn
\ket{S_{k_1}^z S_{k_2}^\pm  }&=&\mp  \frac{1}{2} R(k_1 , k_2 ) \ket{{\cal T}_{k_{1}}^0 {\cal T}_{k_{2}}^\pm  }  . \label{eq:simplified-identity}
\ea
The appearance of the AKLT component in Eq. (\ref{eq:simplified-identity}) is not alarming, but simply reflects the non-orthogonality of the two-magnon or two-triplon state to the AKLT ground state when $k_1 + k_2 =0$ and $\alpha_1 + \alpha_2 =0$. The proper excited state, labeled with the subscript $\perp$ below, is easily constructed by taking away the ground state component from the right side for these cases:
\begin{eqnarray}
|{\cal T}^0_{k_1}{\cal T}^0_{k_2} \rangle_{\perp}  &=&  |{\cal T}^0_{k_1}{\cal T}^0_{k_2} \rangle  + N \delta_{k_1, -k_2} \frac{1+3\cos k_1 }{5+3\cos k_1 }  |A  \rangle  \nn
|{\cal T}^+_{k_1}{\cal T}^-_{k_2} \rangle_{\perp}  &=&  |{\cal T}^+_{k_1}{\cal T}^-_{k_2} \rangle  - \frac{N}{2}  \delta_{k_1, -k_2} \frac{1+3\cos k_1 }{5+3\cos k_1 }
 |A  \rangle . \nn \label{eq:orthogonalized}
\end{eqnarray}
One can check that these new states are orthogonal to the ground state: $\langle A |{\cal T}^0_{k_1}{\cal T}^0_{k_2} \rangle_{\perp}  = \langle A |{\cal T}^+_{k_1}{\cal T}^-_{k_2} \rangle_{\perp} = 0 $.
Orthogonalized two-magnon states are also easily derived, by referring to Eqs. (\ref{eq:simplified-identity}) and (\ref{eq:orthogonalized}).

One can view Eq. (\ref{eq:simplified-identity}) as general statements for the identity of arbitrary two-magnon and two-triplon wave functions of momenta $(k_1, k_2)$. Furthermore, since the two-triplon states have the commuting property $|{\cal T}^{\alpha_1}_{k_1} {\cal T}^{\alpha_2}_{k_2} \rangle = |{\cal T}^{\alpha_2}_{k_2 } {\cal T}^{\alpha_1 }_{k_1} \rangle$ (easily verified from its definition), it follows that the two-magnon wave function on the left must also share the same property: $|S^{\alpha_1}_{k_1} S^{\alpha_2}_{k_2} \rangle = |S^{\alpha_2}_{k_2 } S^{\alpha_1 }_{k_1} \rangle$.

The proof of equivalence of one-magnon and one-triplon wave states in the previous section has been extended to the case of two-magnon and two-triplon wave states. It is obviously tempting to expect the correspondence to continue to multi-magnon and multi-triplon wave functions. In the Appendix \ref{appendix:n} we present the proof of the exact identity between $n$-magnon and $n$-triplon states for the spin-polarized case $\alpha_1 = \cdots  = \alpha_n$. For the more general multi-magnon cases we leave such a statement as a conjecture.

\subsection{Single-mode calculation of the two-triplon energy spectrum}

The magnon and the triplon descriptions proved to be completely equivalent, at least at the one- and two-magnon levels, and it becomes a matter of convenience to choose either description as the excited states. For the calculation of energy it is definitely more convenient to use the triplon representation, because of the equivalence of Schwinger boson representation and the matrix product state representation, and the ability to use the latter scheme to perform the overlap calculation as that of the transfer matrix familiar from one-dimensional statistical mechanics. The procedures are described carefully in the Appendix \ref{appendix:TM}, to which interested readers are referred.

The SMA evaluation of the two-triplon energy involves the expression
\ba \omega_2^{\alpha_1 \alpha_2} ( k_1  , k_2 )  = { \langle
{\cal T}^{\alpha_1}_{k_1} {\cal T}^{\alpha_2}_{k_2}  |H_A |{\cal T}^{\alpha_1}_{k_1} {\cal T}^{\alpha_2}_{k_2} \rangle  \over \langle
{\cal T}^{\alpha_1}_{k_1} {\cal T}^{\alpha_2}_{k_2}  |{\cal T}^{\alpha_1}_{k_1} {\cal T}^{\alpha_2}_{k_2} \rangle } .\ea
To the extent that the triplon-triplon interactions can be neglected, the outcome of such calculation is expected to be a sum of one-triplon energies, $\omega_1 (k_1) + \omega_1 (k_2)$, with $\omega_1 (k)$ found in the one-triplon SMA. Mathematically speaking, such factorized outcome for the energy will be guaranteed provided both the numerator and the denominator could factorize as follows:
\ba \label{eq:factorization} \langle
{\cal T}^{\alpha_1}_{k_1} {\cal T}^{\alpha_2}_{k_2}  |H_A |{\cal T}^{\alpha_1}_{k_1} {\cal T}^{\alpha_2}_{k_2} \rangle  & \stackrel{?}{=} &   \langle
{\cal T}^{\alpha_1}_{k_1}  |H_A |{\cal T}^{\alpha_1}_{k_1}  \rangle   \langle  {\cal T}^{\alpha_2}_{k_2} | {\cal T}^{\alpha_2}_{k_2} \rangle \nn
& + &  \langle
{\cal T}^{\alpha_2}_{k_2}  |H_A |{\cal T}^{\alpha_2}_{k_2}  \rangle   \langle  {\cal T}^{\alpha_1}_{k_1} | {\cal T}^{\alpha_1}_{k_1} \rangle \nn
\langle  {\cal T}^{\alpha_1}_{k_1} {\cal T}^{\alpha_2}_{k_2}  |{\cal T}^{\alpha_1}_{k_1} {\cal T}^{\alpha_2}_{k_2} \rangle  & \stackrel{?}{=} & \langle  {\cal T}^{\alpha_1}_{k_1} |{\cal T}^{\alpha_1}_{k_1} \rangle  \langle   {\cal T}^{\alpha_2}_{k_2}  | {\cal T}^{\alpha_2}_{k_2} \rangle .\ea
As it turns out, both expectations are borne out by explicit calculations of both the numerator and the denominator, for all possible pairs of magnetic quantum numbers $(\alpha_1 ,\alpha_2 )$, leading to an extremely simple, factorized form of the two-triplon energy:
\ba \omega_2^{\alpha_1 \alpha_2}  (k_2 , k_2) = \omega_1 (k_1 ) + \omega_1 (k_2 ).  \label{eq:two-triplon-energy} \ea
The interaction effect, as one might call the correction to the factorized form of energy, is therefore completely absent, at least at the level of SMA.

According to our SMA calculation, the two-magnon spectrum will be 9-fold degenerate without suffering energy-splitting corrections. It is fair to suspect if such lack of interaction effects and level repulsion among the degenerate states might be an artifact of the SMA. Interestingly, the authors of Refs.~\cite{haegeman12,ng14} have found an almost complete degeneracy of all two-magnon channels with total spins $S=2,1,0$ - 9-fold degeneracy in total - in the numerical evaluation based on the Heisenberg spin chain. The model is different from the AKLT Hamiltonian, but still both of their ground states are known to belong to the same Haldane phase. If the correspondence should extend to excited state properties as well, the strict 9-fold degeneracy predicted by our SMA two-magnon energy calculation, valid for the AKLT model, is in line with those numerical observations made on the Heisenberg model.

Let us now discuss the details of SMA calculation for the case $\alpha_1 = \alpha_2 = +$. Most of the details of the overlap derivation are delegated to the two Appendices \ref{appendix:denominator} and \ref{appendix:numerator}. It can be shown,
\begin{widetext}
\ba
\langle {\cal T}_{k_{1}}^{+}{\cal T}_{k_{2}}^{+}|{\cal T}_{k_{1}}^{+}{\cal T}_{k_{2}}^{+} \rangle
= \langle {\cal T}_{k_{1}}^{-}{\cal T}_{k_{2}}^{-}|{\cal T}_{k_{1}}^{-}{\cal T}_{k_{2}}^{-} \rangle &=& \frac{ 4(1 + \delta_{k_1,k_2})}{(5+3\cos{k_1})(5+3\cos{k_2})} 3^N N^2. \nn
\langle {\cal T}_{k_{1}}^{+}{\cal T}_{k_{2}}^{+}|H_A | {\cal T}_{k_{1}}^{+}{\cal T}_{k_{2}}^{+} \rangle = \langle {\cal T}_{k_{1}}^{-}{\cal T}_{k_{2}}^{-}|H_A | {\cal T}_{k_{1}}^{-}{\cal T}_{k_{2}}^{-} \rangle   &=& {20 \over 27} \left(\frac{1}{5 + 3 \cos k_1 } + \frac{1}{5 + 3 \cos k_2 }\right)(1+ \delta_{k_1,k_2})3^N N^2 . \label{eq:overlap1} \ea
On taking their ratio, we recover the two-triplon energy in the factorized form: $\omega_2^{++} (k_1 , k_2 ) = \omega_1 (k_1 ) + \omega_1 (k_2 )$. Other cases follow suit in a similar fashion:
\ba
\langle {\cal T}_{k_1}^{+} {\cal T}_{k_2}^{0} | {\cal T}_{k_1}^{+} {\cal T}_{k_2}^{0}\rangle = \langle{\cal T}_{k_1}^{-} {\cal T}_{k_2}^{0} | {\cal T}_{k_1}^{-} {\cal T}_{k_2}^{0}\rangle &=&\frac{8}{(5+3 \cos k_1) (5+3 \cos k_2)}3^N N^2 \nn
\langle {\cal T}_{k_1}^{+} {\cal T}_{k_2}^{0} | H_A | {\cal T}_{k_1}^{+} {\cal T}_{k_2}^{0}\rangle = \langle{\cal T}_{k_1}^{-} {\cal T}_{k_2}^{0} |H_A | {\cal T}_{k_1}^{-} {\cal T}_{k_2}^{0}\rangle &=&{ 40 \over 27} \left(\frac{1}{5 + 3\cos k_1 } + \frac{1}{5 + 3 \cos k_2}\right) 3^N N^2 . \ea
In the remaining cases $(\alpha_1, \alpha_2) = (0,0)$, $(+,-)$, $(-,+)$ one needs to use the orthogonalized excited states given in (\ref{eq:orthogonalized}) to compute the overlaps:
\ba
\langle{\cal T}_{k_{1}}^{0}{\cal T}_{k_{2}}^{0}|{\cal T}_{k_{1}}^{0}{\cal T}_{k_{2}}^{0} \rangle &=&
\frac{ 16(1 + \delta_{k_1,k_2})}{(5+ 3 \cos k_1) (5+ 3 \cos k_2)} 3^N N^2 \nn
\langle{\cal T}_{k_{1}}^{0}{\cal T}_{k_{2}}^{0}|H_A | {\cal T}_{k_{1}}^{0}{\cal T}_{k_{2}}^{0} \rangle &=& { 80 \over 27}(1 + \delta_{k_1,k_2}) \left(\frac{1}{5 + 3\cos k_1 } + \frac{1}{5 + 3 \cos k_2}\right) 3^N N^2 \nn
\langle {\cal T}_{k_1}^{+} {\cal T}_{k_2}^{-} | {\cal T}_{k_1}^{+} {\cal T}_{k_2}^{-}\rangle  &=&
\frac{ 4 }{(5+ 3 \cos k_1) (5+ 3 \cos k_2)} 3^N N^2 \nn
\langle {\cal T}_{k_1}^{+} {\cal T}_{k_2}^{-} |H_A | {\cal T}_{k_1}^{+} {\cal T}_{k_2}^{-}\rangle  &=& { 20 \over 27} \left(\frac{1}{5 + 3\cos k_1 } + \frac{1}{5 + 3 \cos k_2}\right) 3^N N^2 .
\ea
\end{widetext}
The subscripts $\perp$ are dropped here for notational simplicity. In all cases, the factorized two-triplon energy form in Eq. (\ref{eq:two-triplon-energy}) follows.

All of these is not to say that the two-triplon state is the eigenstate of the AKLT Hamiltonian. Even the one-triplon state is not the eigenstate, as the action of $H_A$ on the one-triplon state is given by
\ba
&& H_A\ket{{\cal T}_{k}^1} = \left( 1+\frac{2}{3}\cos k \right)\ket{{\cal T}_{k}^1}\nn
&& ~~~  +\frac{1}{6}\sum_ie^{i k x_i}(\ket{{\cal T}_{i-1}^{1}{\cal T}_{i+1}^{0}}-\ket{{\cal T}_{i-1}^{0}{\cal T}_{i+1}^{1}}) . \label{eq:H-on-one-triplon} \ea
The action of $H_A$ on the one-triplon wave generates additional two-triplon states shown in the second line. The action of $H_A$ on the two-triplon wave is, for instance,
\ba
&& H_A\ket{{\cal T}_{k_1}^1{\cal T}_{k_2}^1}
=\left( 2 +  \frac{2}{3}\cos k_1+\frac{2}{3}\cos k_2 \right)\ket{{\cal T}_{k_1}^1{\cal T}_{k_2}^1}
\nn
&&+\frac{1}{6} \sum_{i,j;i\neq j} (e^{ik_1x_i+ik_2x_j} + e^{ik_1x_j+ik_2x_i}) \ket{{\cal T}_{i-1}^{1}{\cal T}_{i+1}^{0}{\cal T}_{j}^{1}} \nn
&&-\frac{1}{6} \sum_{i,j;i\neq j} (e^{ik_1x_i+ik_2x_j} + e^{ik_1x_j+ik_2x_i}) \ket{{\cal T}_{i-1}^{0}{\cal T}_{i+1}^{1}{\cal T}_{j}^{1}} . \nn \label{eq:H-on-two-triplon}
\ea
In both instances, the factorized energy form is recovered on taking the overlap with the ket. The same thing happens with all other two-triplon wave states.

\section{Construction of effective Hamiltonian}
\label{sec:effective-H}
The one- and two-triplon wave states we have constructed are not the eigenstates of the AKLT Hamiltonian, but they may still serve as good basis states spanning the low-energy excited modes. Indeed, systematic construction of general $n$-magnon wave states as the basis states (not necessarily eigenstates) which span the excitation spectrum in a ferromagnet was at the heart of Dyson's program~\cite{dyson56}. Here, due to technical challenges, we are limited to including only the one- and two-triplon wave basis states in constructing the low-energy manifold, but given that the $n$-triplon waves are likely to have energies about $n$ times that of a one-triplon, this may not be a harsh restriction. The program we carry out here is, first of all, the construction of properly orthogonalized one- and two-triplon wave states, and secondly the construction of effective Hamiltonian within such subspace.

First of all, the nine two-triplon states $| {\cal T}^{\alpha_1}_{k_1} {\cal T}^{\alpha_2}_{k_2} \rangle$ are further classified according to their total angular momentum being 2, 1, and 0. Invoking the usual Clebsch-Gordon argument, we list the five quintuplet as
\ba | {\cal T}^{2,2}_{k_1, k_2} \rangle & = & \ket{\overline{{\cal T}_{k_1}^1{\cal T}_{k_2}^1}}\nn
| T^{2,1}_{k_1, k_2} \rangle & = & \frac{1}{\sqrt{2}} \left( \ket{\overline{{\cal T}_{k_1}^1{\cal T}_{k_2}^0}} + \ket{\overline{{\cal T}_{k_1}^0{\cal T}_{k_2}^1}} \right) \nn
| {\cal T}^{2,0}_{k_1, k_2} \rangle & = & \frac{1}{\sqrt{6}} \left( \ket{\overline{{\cal T}_{k_1}^{1}{\cal T}_{k_2}^{-1}}}_{\perp}+  \ket{\overline{{\cal T}_{k_1}^{-1}{\cal T}_{k_2}^{1}}}_{\perp}+2 \ket{\overline{{\cal T}_{k_1}^{0}{\cal T}_{k_2}^{0}}}_{\perp} \right) \nn
| {\cal T}^{2,-1}_{k_1, k_2} \rangle & = & \frac{1}{\sqrt{2}} \left( \ket{\overline{{\cal T}_{k_1}^{-1}{\cal T}_{k_2}^0}}+\ket{\overline{{\cal T}_{k_1}^1{\cal T}_{k_2}^{-1}}} \right) \nn
| {\cal T}^{2,-2}_{k_1, k_2} \rangle & = & \ket{\overline{{\cal T}_{k_1}^{-1}{\cal T}_{k_2}^{-1}}}
\ea
where the overline means a normalized state. The subscript $\perp$ was introduced earlier in Eq. (\ref{eq:orthogonalized}) to define the orthogonalized two-triplon wave state having zero overlap with the one-triplon state. The singlet combination is given by
\ba | {\cal T}^{0,0}_{k_1, k_2} \rangle = \frac{1}{\sqrt{3}} \left( \ket{\overline{{\cal T}_{k_1}^{-1}{\cal T}_{k_2}^{1}}}_{\perp}+ \ket{\overline{{\cal T}_{k_1}^{1}{\cal T}_{k_2}^{-1}}}_{\perp} - \ket{\overline{{\cal T}_{k_1}^{0}{\cal T}_{k_2}^{0}}}_{\perp} \right) . \nn
\ea
To verify the angular momentum properties of the above states, one can first show that the two-magnon waves can be organized as total angular momentum eigenstates with $S=0,1,2$, and subsequently invoking the triplon-magnon equivalence  via  Eq.~(\ref{eq:correspondence}) to argue the same for the two-triplon waves. The final conclusion is, after all, exactly what one expects from the Clebsch-Gordon algebra. 

Neither the quintuplet nor the singlet should have any mixing with the triplet state of one-triplons $| T^\alpha_k \rangle$, by virtue of angular momentum mismatch. Therefore, the only kind of one- and two-triplon mixing one needs to worry about is with the triplet combination of the two-triplons, given by
\ba
| {\cal T}^{1,1}_{k_1, k_2} \rangle & = & \frac{1}{\sqrt{2}} \left( |\overline{{\cal T}^1_{k_1} {\cal T}^0_{k_2}} \rangle -  |\overline{{\cal T}^0_{k_1} {\cal T}^1_{k_2}} \rangle \right)  \nn
| {\cal T}^{1,0}_{k_1, k_2} \rangle & = & \frac{1}{\sqrt{2}} \left( |\overline{{\cal T}^1_{k_1} {\cal T}^{-1}_{k_2}} \rangle_{\perp} - |\overline{{\cal T}^{-1}_{k_1} {\cal T}^1_{k_2}} \rangle_{\perp} \right) \nn
| {\cal T}^{1,-1}_{k_1, k_2} \rangle & = & \frac{1}{\sqrt{2}} \left( |\overline{{\cal T}^0_{k_1} {\cal T}^{-1}_{k_2}} \rangle -  |\overline{{\cal T}^{-1}_{k_1} {\cal T}^0_{k_2}} \rangle \right) . \label{eq:two-t-triplet} \ea
The only nonzero matrix elements between one- and two-triplon sector is then
$\langle \overline{{\cal T}^\alpha_{k_1+ k_2}} | {\cal T}^{1,\alpha}_{k_1, k_2} \rangle$ and $\langle \overline{{\cal T}^\alpha_{k_1+ k_2}} | H | {\cal T}^{1,\alpha}_{k_1, k_2} \rangle$,
both of which can be worked out through explicit calculations. The results are
\begin{widetext}
\ba
\langle \overline{ {\cal T}^\alpha_{k_1+ k_2} }  | {\cal T}^{1,\alpha}_{k_1, k_2} \rangle&=& - \frac{3i }{2\sqrt{N}} \frac{ 3 \sin (k_1-k_2)+13 \sin k_1-13 \sin k_2 }{ \sqrt{2(5 + 3 \cos k_1 ) (5 + 3 \cos k_2 ) (5 + 3 \cos (k_1+k_2) )}}\nn
\langle \overline{ {\cal T}^\alpha_{k_1+ k_2} }  | H | {\cal T}^{1,\alpha}_{k_1, k_2} \rangle&=& - \frac{1}{\sqrt{N}}  {5i \over 6}   (3 \sin (k_1-k_2)+5 \sin k_1-5 \sin k_2)  \frac{ \sqrt{5 + 3 \cos (k_1+k_2)} }{ \sqrt{2( 5+ 3 \cos k_1 ) (5 + 3 \cos k_2 )}} . \label{eq:1-2-overlap}
\ea
\end{widetext}
The overlap factors are of order $1/\sqrt{N}$ and vanishing in the thermodynamic limit, but it does not imply that they make negligible contribution to physical quantities, as we will shortly show.

There is one more piece of preparatory work required before one can declare the completion of the effective Hamiltonian construction within the one- and two-triplon subspace. The two-triplon triplet state constructed in Eq. (\ref{eq:two-t-triplet}) is not properly orthogonalized with respect to the singlet-triplon states while, generically, the effective Hamiltonian must be constructed in the basis of orthogonal states. To that end, we implement the Gram-Schmidt orthogonalization procedure and introduce the properly orthogonalized two-triplon triplet basis states
\ba
| {\cal T}^{1,\alpha}_{k_1, k_2} \rangle_\perp & = & | {\cal T}^{1,\alpha}_{k_1, k_2} \rangle -  \langle \overline{ {\cal T}^\alpha_{k_1+ k_2} }  | {\cal T}^{1,\alpha}_{k_1, k_2} \rangle
|\overline{{\cal T}^\alpha_{k_1 + k_2}} \rangle   . \label{eq:two-t-triplet-ortho}\nn
\ea
The required matrix element $\langle \overline{ {\cal T}^\alpha_{k_1+ k_2} }  | {\cal T}^{1,\alpha}_{k_1, k_2} \rangle$ is already given in Eq. (\ref{eq:1-2-overlap}). In this basis the Hamiltonian matrix element reads, instead of the second line of Eq. (\ref{eq:1-2-overlap}), $\langle \overline{ {\cal T}^\alpha_{k_1+ k_2} }  | H | {\cal T}^{1,\alpha}_{k_1, k_2} \rangle_\perp = -  i f(k_1 , k_2)/\sqrt{N} $ with the form factor
\begin{widetext}
\ba
f(k_1, k_2 ) =  {5\sqrt{2}  \over 18}  (\sin{k_1} - \sin{k_2} + 3\sin{(k_1-k_2)}) { \sqrt{5+3\cos{(k_1+k_2)}} \over \sqrt{(5+3\cos{k_1})(5+3\cos{k_2})} } .   \nn
\ea
Finally, the matrix elements within the two-triplon triplet sector is
\ba
\langle  {\cal T}^{1,\alpha}_{k_3 ,  k_4}   |_\perp H | {\cal T}^{1,\alpha}_{k_1, k_2} \rangle_\perp = (\omega_1(k_1) + \omega_1(k_2))  (\delta_{k_1,k_3} \delta_{k_2,k_4} - \delta_{k_1,k_4} \delta_{k_2,k_3})  .
\ea
\end{widetext}
There is a sub-leading factor of order $1/N$ that takes place when $k_1 \neq k_3$, which can be ignored because it falls with a faster power than $1/\sqrt{N}$ and does not contribute in the physical quantities. The relative minus sign between the two delta functions on the right side can be understood from the antisymmetry of the wave function: $| {\cal T}^{1,\alpha}_{k_1, k_2} \rangle = - | {\cal T}^{1,\alpha}_{k_2, k_1} \rangle$.

\begin{figure}[!ht]
\includegraphics[width=5cm,height=5cm]{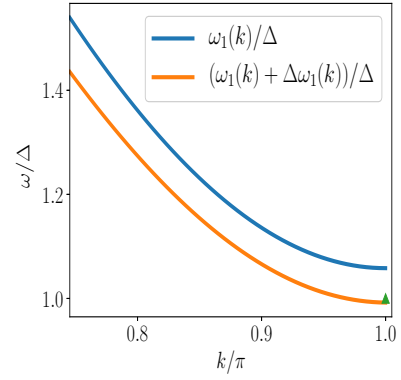}
\caption{SMA energy with perturbative second-order correction. The vertical axis is renormalized by the exact energy gap at $k=\pi$, which is designated as $\Delta$ and shown as a triangle at $k=\pi$.}
\label{fig:fig1}
\end{figure}

In the end, we have a quite simple structure of the effective Hamiltonian consisting of the following matrix elements:
\ba \langle {\cal T}^\alpha_k |H |{\cal T}^\alpha_k  \rangle &=& \omega_1(k) \nn
\langle {\cal T}^\alpha_{k_1 + k_2} | H |{\cal T}^{1,\alpha}_{k_1, k_2 } \rangle_\perp &=& -  i f(k_1 , k_2 )/\sqrt{N}  \nn
\langle  {\cal T}^{1,\alpha}_{k_1 ,  k_2}   |_\perp H | {\cal T}^{1,\alpha}_{k_1, k_2} \rangle_\perp   &= & \omega_1(k_1) + \omega_1(k_2).  \ea
We no longer place a bar to indicate normalized states; all one- and two-triplon states are by now assumed properly normalized and orthogonalized. An immediate consequence of the effective Hamiltonian is the level repulsion between the ``bare" one-triplon energy $\omega_1 (k)$ and the ``bare" two-triplon energy $\omega_2 (k_1 , k_2 ) = \omega_1 (k_1 ) + \omega_1 (k_2)$ worked out in the SMA calculation of the previous section. The one-triplon energy shift is  given by
\ba \Delta \omega_1 (k) & = & - \sum_{k_1 + k_2 = k }{  | \langle  {\cal T}^{1, \alpha}_{k_1, k_2} |H |{\cal T}^\alpha_k \rangle |^2 \over \omega_1 (k_1 )  + \omega_1 (k_2 ) -   \omega_1 (k) } \nn
& = & - {1\over N} \sum_{k_1 + k_2 = k }{  [  f(k_1 , k_2 ) ]^2 \over \omega_1 (k_1 )  + \omega_1 (k_2 ) -   \omega_1 (k) } \nn
& \rightarrow  & -  \int_{- \pi + {k \over 2}}^{\pi-{k \over 2} } {dq\over 2\pi}  {  [  f(k/2 \!+\! q , k/2 \!-\! q) ]^2 \over \omega_1 (k/2 \!+\! q )  \!+\! \omega_1 (k/2 \!-\! q ) \!- \!  \omega_1 (k) }    .  \nn \ea
The correction formula $\Delta \omega_1 (k)$ is valid as long as we are sufficiently removed from $k=\pi/2$, where the crossing of the single- and two-triplon bare energies occurs and the denominator in the perturbative formula vanishes.
The one-triplon states are best defined near $k=\pi$ after all, and this is where the perturbative scheme should work most excellently. The bare and the corrected single-triplon energies are shown in Fig.~\ref{fig:fig1} near $k=\pi$. The bare energy at $k=\pi$ is 0.370, compared to the exact numerical value 0.350~\cite{mikeska95,Wei13}. After the perturbative correction, the energy becomes 0.347, much closer to the exact one.

On the other hand, the correction in the two-triplon energy is
\ba \Delta \omega_2 (k_1 , k_2) \!=\! {1\over N} { [f(k_1 , k_2 ) ]^2 \over \omega_2 (k_1 , k_2 ) \!- \!\omega_1 (k_1 ) \!-\!\omega_1 ( k_2 ) } . \ea
This factor is vanishing in the thermodynamic limit, and does not lift the 9-fold degeneracy of the two-triplon SMA energy.
What happens is obviously that the one-triplon state couples to many two-triplon states with a small amplitude $\sim 1/\sqrt{N}$, with a net effect that remains finite, while the two-triplon state couples to only one state in the one-triplon sector with amplitude $1/\sqrt{N}$ and negligible effects. In conclusion, there is no lifting of either the one-triplon energy or the two-triplon energy worked out from SMA. The former conclusion is of course expected from symmetry, but the protection of the 9-fold two-triplon energy is non-trivial.

In regard to the wave function, the correction to the two-triplon wave function can be neglected in the large-$N$ limit, but not the one on the one-triplon wave function. We have, after the hybridization, the modified one-triplon wave function
\ba |\overline{{\cal T}^\alpha_k} \rangle \rightarrow |\overline{{\cal T}^\alpha_k} \rangle - {i \over \sqrt{N} } \sum_{k_1 + k_2 = k} f(k_1 , k_2 ) |{\cal T}^{1,\alpha}_{k_1, k_2 } \rangle. \ea
One can make use of the low-energy space of one- and two-triplons and the effective Hamiltonian constructed within such space to explore various dynamic and thermodynamic responses. Such task will be taken up in a subsequent calculation, as the bulk of calculation one has to digest through in this publication alone is already very heavy.

\section{Summary}
\label{sec:discussion}
Previously known equivalence of one-magnon and one-triplon excited states of the AKLT chain model has been extended to that of two-magnon and two-triplon excited states in this work. In some cases, exact correspondence of $n$-magnon and $n$-triplon states has been established as well.

By exploiting the equivalence of the Schwinger boson and matrix product state representations of the spin-1 chain state, and the subsequent transfer matrix formalism, we were able to compute the two-triplon energies in the single-mode approximation. The two-triplon energy is proven to break down exactly as the sum of two, one-triplon energies, without corrections. Although the two-triplon states are not themselves the exact eigenstates of the AKLT Hamiltonian, the complete lack of interaction effects among triplon waves at respective momenta $k_1$ and $k_2$ in the SMA calculation is interesting. It reminds one of several numerical studies showing almost perfect nine-fold degeneracy of two-magnon excited states in the Heisenberg spin Hamiltonian. Furthermore, we constructed an effective Hamiltonian within the one- and two-triplon subspace and computed perturbative corrections to  the energy and the one-triplon states. The value of the corrected energy gap was within $1\%$ of the actual value, reaching the same accuracy as that obtained in Ref. \cite{mikeska95} using  solitonic ansatz with smeared out domain walls. Our effective Hamiltonian scheme can serve purposes beyond that of  calculating energy correction, though, and opens up the possibility to compute wide range of dynamical properties. 

The program initiated by Dyson provided a useful basis for treating the many-body magnon dynamics and thermodynamics of ferromagnets. It was, in a sense, a program carried out in a ``non-interacting" picture where the ground state is a simple product state (all spins pointing in the same direction), and the $n$-magnon states constructed on top of it were approximately orthogonal. Here we have dealt with an analogous program, trying to build orthogonal sets of excited states on top of a ``correlated" ground state, {\it i. e.} the AKLT state. Surprisingly, despite the complications brought by the correlated nature of the ground state, construction of mutually orthogonal excited states was proven to be possible. Effective Hamiltonian constructed in the low-energy magnon (triplon) space revealed only weak interaction matrix elements between one- and two-magnon (one- and two-triplon) basis states. Extending the scheme to higher-magnon states remains a challenge. Application of the scheme developed in this paper to the calculation of various dynamic and thermodynamic quantities, and comparison of such results to numerical simulations, remains as future work. 

\acknowledgments
J. H. H. was supported by Samsung Science and Technology Foundation under Project Number SSTF-BA1701-07.

\appendix

\section{Overlap Calculation of MPS States}
\label{appendix:TM}

The Schwinger boson representation of the AKLT ground state $|A\rangle$ allows an equivalent matrix product state (MPS) expression~\cite{zittartz92}. Schematically, the correspondence can be expressed as
\ba && |A\rangle = [ \prod_i {\cal S}_{i,i+1} ]  |v\rangle \leftrightarrow \nn
&& (-1)^N \sum_{\{s\} }{\rm Tr} [M_0^{(s_{1})}M_0^{(s_{2})}\cdots M_0^{(s_{N})}] |\{s \} \rangle . \label{eq:AKLT-MPS} \ea
The summation over all possible spin orientations $\{s\} = \{ s_1, \cdots, s_N \}$ is performed in the second line. The $2\times2$ matrices defined here for each spin orientation $s=1$, $0$, $-1$ are given by
\begin{align}
{M}_0^{(1)}&= \begin{pmatrix}
0 & -\sqrt{2}\\
0 & 0\end{pmatrix} &
M_0^{(0)}&=\begin{pmatrix}
1 & 0\\
0 & -1\end{pmatrix} &
M_0^{(-1)} &= \begin{pmatrix}
0 & 0\\
\sqrt{2} & 0\end{pmatrix} .\nn
\end{align}
Such one-to-one correspondence extends well beyond the AKLT state, and in fact covers an {\it arbitrary} state in which one replaces the Schwinger boson singlet ${\cal S}_{i,i+1}$ by one of the triplets ${\cal T}_{i,i+1}^\alpha$.

First introduce an additional set of matrices
\begin{align}\label{eq:sol}
M_{1}^{(1)} &= \begin{pmatrix}
0 & 0\\
0 & 0\end{pmatrix} &
M_{1}^{(0)} &= \begin{pmatrix}
0 & 1\\
0 & 0\end{pmatrix} &
M_{1}^{(-1)} &= \begin{pmatrix}
0 & 0\\
0 & \sqrt{2}\end{pmatrix}\nn
M_{2}^{(1)} &= \begin{pmatrix}
0 & \sqrt{2}\\
0 & 0\end{pmatrix}\nonumber&
M_{2}^{(0)} &= \begin{pmatrix}
1 & 0\\
0 & 1\end{pmatrix} &
M_{2}^{(-1)} &= \begin{pmatrix}
0 & 0\\
\sqrt{2} & 0\end{pmatrix}\\
M_{3}^{(1)} &= \begin{pmatrix}
\sqrt{2} & 0\\
0 & 0\end{pmatrix} &
M_{3}^{(0)} &= \begin{pmatrix}
0 & 0\\
1 & 0\end{pmatrix} &
M_{3}^{(-1)} &= \begin{pmatrix}
0 & 0\\
0 & 0\end{pmatrix} .
\end{align}
The upper index $s$ still refers to the spin orientation of the basis state. The three lower indices $1$, $2$, $3$ correspond to the $\alpha=-1$, $0$, $1$ components of the triplet operator ${\cal T}^\alpha_{i,i+1}$. Whenever a particular singlet ${\cal S}_{i,i+1}$ in the AKLT state is replaced by the triplet ${\cal T}_{i,i+1}^\alpha$, one replaces $M_0^{(s_i)}$ by $M_{\alpha +2}^{(s_i)}$, where $\alpha+2$ runs through $1$, $2$, $3$ as in (\ref{eq:sol}). For instance, a one-triplon state has the MPS representation
\ba \label{eq:triplon-MPS}
\ket{{\cal T}_i^{\alpha}}
&=&(-1)^{N-1}\sum_{\{s\}}\textrm{Tr}[M_0^{(s_{1})}\cdots M_{\alpha+2}^{(s_{i})}\cdots M_0^{(s_{N})}]\ket{\{ s\}} .\nn
\ea
A nice way to account for the sign factors in (\ref{eq:AKLT-MPS}) and (\ref{eq:triplon-MPS}) is to remember the replacement rule
\ba\label{eq:rep-rule}
[ {\cal S}_{i,i+1} ]^\dag \rightarrow - M^{(s)}_0 ~~~
[{\cal T}_{i,i+1}^{\alpha}]^{\dagger} \rightarrow M^{(s)}_{\alpha+2}  .
\ea

Now that every Schwinger boson state has an equivalent MPS representation, their overlaps can also be evaluated by invoking their MPS forms. For two arbitrary MPS states $|\psi\rangle$ and $|\psi'\rangle$, their overlap is
\ba
\ket{\psi} &=& \sum_{\{s\}}\textrm{Tr}[A_1^{(s_{1})}A_2^{(s_{2})}\cdots A_N^{(s_{N})}]\ket{\{s\}}\nn
\ket{\psi'} &=& \sum_{\{s\}}\textrm{Tr}[B_1^{(s_{1})}B_2^{(s_{2})}\cdots B_N^{(s_{N})}]\ket{\{s\}}\nn
\braket{\psi'|\psi} &=& \sum_{\{s\}}\textrm{Tr}[B_1^{(s_{1})}B_2^{(s_{2})}\cdots B_N^{(s_{N})}]\textrm{Tr}[A_1^{(s_{1})}A_2^{(s_{2})}\cdots A_N^{(s_{N})}]\nn
\ea
where $A_i$, $B_i=M_0$, $M_1$, $M_2$, $M_3$.

Employing some matrix identities
\ba
\textrm{Tr}[A]\textrm{Tr}[B] &=& \textrm{Tr}[A\otimes B]\nn
ABC\otimes DEF &=& (A\otimes D)(B\otimes E)(C\otimes F)
\ea
one can rewrite the overlap
\begin{widetext}
\ba
\braket{\psi'|\psi} &=&\sum_{\{s\}} \textrm{Tr}[(B_1^{(s_{1})}B_2^{(s_{2})}\cdots B_N^{(s_{N})})\otimes(A_1^{(s_{1})}A_2^{(s_{2})}\cdots A_N^{(s_{N})})]\nn
&=&\sum_{\{s\}}\textrm{Tr}[(B_1^{(s_{1})}\otimes A_1^{(s_{1})})(B_2^{(s_{2})}\otimes A_2^{(s_{2})})\cdots (B_N^{(s_{N})}\otimes A_N^{(s_{N})})]\nn
&=&\textrm{Tr}[(\sum_{s_1}B_1^{(s_{1})}\otimes A_1^{(s_{1})})(\sum_{s_2}B_1^{(s_{2})}\otimes A_1^{(s_{2})})\cdots(\sum_{s_N}B_1^{(s_{N})}\otimes A_1^{(s_{N})})].
\ea
\end{widetext}
Since there are 4 possibilities for each matrix, there are 16 cases of the direct product $\sum_{s_i}B_i^{(s_{i})}\otimes A_i^{(s_{i})}$ in all. Define a matrix $M_{ij}$ and overlap MPS form of $\braket{\psi'|\psi}$
\ba
M_{ij} &=& \sum_{s}M_{i}^{(s)}\otimes M_{j}^{(s)}\nn
\braket{\psi'|\psi} &=& \textrm{Tr}[M_{i_1 j_1} M_{i_2 j_2} \cdots M_{i_N j_N} ]
\ea
The sixteen $M_{ij}$ matrices are given by
%
\begin{align}\label{eq:M_ij}
M_{00} &= \begin{pmatrix}
1 & 0 & 0 & 2\\
0 & -1 & 0 & 0\\
0 & 0 & -1 & 0\\
2 & 0 & 0 & 1\end{pmatrix}&
M_{01} &= \begin{pmatrix}
0 & 1 & 0 & 0\\
0 & 0 & 0 & 0\\
0 & 0 & 0 & -1\\
0 & 2 & 0 & 0\end{pmatrix}\nn
M_{02} &= \begin{pmatrix}
1 & 0 & 0 & -2\\
0 & 1 & 0 & 0\\
0 & 0 & -1 & 0\\
2 & 0 & 0 & -1\end{pmatrix}&
M_{03} &= \begin{pmatrix}
0 & 0 & -2 & 0\\
1 & 0 & 0 & 0\\
0 & 0 & 0 & 0\\
0 & 0 & -1 & 0\end{pmatrix}\nn
M_{10} &= \begin{pmatrix}
0 & 0 & 1 & 0\\
0 & 0 & 0 & -1\\
0 & 0 & 0 & 0\\
0 & 0 & 2 & 0\end{pmatrix}&
M_{11} &= \begin{pmatrix}
0 & 0 & 0 & 1\\
0 & 0 & 0 & 0\\
0 & 0 & 0 & 0\\
0 & 0 & 0 & 2\end{pmatrix}\nn
M_{12} &= \begin{pmatrix}
0 & 0 & 1 & 0\\
0 & 0 & 0 & 1\\
0 & 0 & 0 & 0\\
0 & 0 & 2 & 0\end{pmatrix}&
M_{13} &= \begin{pmatrix}
0 & 0 & 0 & 0\\
0 & 0 & 1 & 0\\
0 & 0 & 0 & 0\\
0 & 0 & 0 & 0\end{pmatrix}\nn
M_{20} &= \begin{pmatrix}
1 & 0 & 0 & -2\\
0 & -1 & 0 & 0\\
0 & 0 & 1 & 0\\
2 & 0 & 0 & -1\end{pmatrix}&
M_{21} &= \begin{pmatrix}
0 & 1 & 0 & 0\\
0 & 0 & 0 & 0\\
0 & 0 & 0 & 1\\
0 & 2 & 0 & 0\end{pmatrix}\nn
M_{22} &= \begin{pmatrix}
1 & 0 & 0 & 2\\
0 & 1 & 0 & 0\\
0 & 0 & 1 & 0\\
2 & 0 & 0 & 1\end{pmatrix}&
M_{23} &= \begin{pmatrix}
0 & 0 & 2 & 0\\
1 & 0 & 0 & 0\\
0 & 0 & 0 & 0\\
0 & 0 & 1 & 0\end{pmatrix}\nn
M_{30} &= \begin{pmatrix}
0 & -2 & 0 & 0\\
0 & 0 & 0 & 0\\
1 & 0 & 0 & 0\\
0 & -1 & 0 & 0\end{pmatrix}&
M_{31} &= \begin{pmatrix}
0 & 0 & 0 & 0\\
0 & 0 & 0 & 0\\
0 & 1 & 0 & 0\\
0 & 0 & 0 & 0\end{pmatrix}\nn
M_{32} &= \begin{pmatrix}
0 & 2 & 0 & 0\\
0 & 0 & 0 & 0\\
1 & 0 & 0 & 0\\
0 & 1 & 0 & 0\end{pmatrix}&
M_{33} &= \begin{pmatrix}
2 & 0 & 0 & 0\\
0 & 0 & 0 & 0\\
0 & 0 & 0 & 0\\
1 & 0 & 0 & 0\end{pmatrix} .
\end{align}
%

Now the overlap calculation $\langle \psi'| \psi\rangle$ reduces to taking the trace of a long product of matrices. Such tasks are performed in statistical mechanics by the transfer matrix method, and we can adopt the same strategy here to compute overlaps. Starting with the simplest case of $\langle A | A \rangle$, we find
\ba\label{eq:OA}
\ket{A} &=& (-1)^N \sum_{\{s\}} \textrm{Tr}[M_0^{(s_1)}\cdots M_0^{(s_N)}]\nn
\braket{A|A}&=&\textrm{Tr}[M^N_{00}] =3^N .
\ea
First of all, the $M_{00}$ matrix can be diagonalized through the unitary rotation $P$.
\begin{align}
P^{-1}M_{00}P&= \begin{pmatrix}
3 & 0 & 0 & 0\\
0 & -1 & 0 & 0\\
0 & 0& -1 & 0\\
0 & 0 & 0& -1\end{pmatrix} &
P &=\begin{pmatrix}
-1 & -1 & 0 & 0\\
0 & 0 & 0 & 1\\
0 & 0& 1 & 0\\
-1 & 1 & 0& 0\end{pmatrix}
\end{align}
Among the four diagonal values, 3 becomes $3^N$ after the matrix multiplication and dominates over all other factors of order $(-1)^N$, hence $\langle A | A \rangle = 3^N$ in the $N\rightarrow \infty$ limit.

In the case of $\braket{{\cal T}_j^{\alpha}|{\cal T}_i^{\alpha}}$ overlap, one replaces $M_0^{(s_i)}$ and $M_0^{(s_j)}$ by $M_{\alpha+2}^{(s_i)}$ and $M_{\alpha+2}^{(s_j)}$, respectively, in the MPS form. We have the transfer matrix form of the overlap when $i>j$
\ba
\braket{{\cal T}_j^{\alpha}|{\cal T}_i^{\alpha}}&=& \textrm{Tr}[M_{0,\alpha+2}M_{00}^{i-j-1}M_{\alpha+2,0}M_{00}^{N-1-i+j}]\nn
&=&\frac{1+\delta_{\alpha,0}}{2}\left((-1)^N(-\frac{1}{3})^{-i+j}+3^N(-\frac{1}{3})^{i-j}\right).\nn
\ea
This leads to Eq. (\ref{eq:1TO}).

The overlap of two-triplon state with another two-triplon state $\braket{{\cal T}_{k}^{\alpha}{\cal T}_{m}^{\beta}|{\cal T}_{i}^{\gamma}{\cal T}_{j}^{\delta}}$ which appears a lot in Appendix C and D has the MPS form
\begin{widetext}
\begin{equation}
\braket{{\cal T}_{k}^{\alpha}{\cal T}_{m}^{\beta}|{\cal T}_{i}^{\gamma}{\cal T}_{j}^{\delta}} = \textrm{Tr}[M_{0,\alpha+2}M_{00}^{i-k-1}M_{\gamma+2,0}M_{00}^{m-i-1}M_{0,\beta+2}M_{00}^{j-m-1}M_{\delta+2,0}M_{00}^{N-j+k-1}]
\end{equation}
\end{widetext}
for the $k<i<m<j$ case. Again, one can employ the transfer matrix method to compute the overlap in the thermodynamic limit.

\section{Equivalence of $n$-Magnon and $n$-Triplon Excitations}
\label{appendix:n}

We can prove the following equivalence of $n$-magnon $\ket{S_{k_1}^{+}S_{k_2}^{+}\cdots S_{k_n}^{+}}$ and the $n$-triplon $\ket{{\cal T}_{k_1}^{1}{\cal T}_{k_2}^{1}\cdots {\cal T}_{k_n}^{1}}$ wave functions by the method of induction:
\ba
\ket{S_{k_1}^{+}S_{k_2}^{+}\cdots S_{k_n}^{+}}= \left[ \Pi_{j=1}^n P_j  \right] \ket{{\cal T}_{k_1}^{1}{\cal T}_{k_2}^{1}\cdots {\cal T}_{k_n}^{1}} . \nn
\ea
The phase factor $P_j = e^{ik_j} - 1$ is introduced as abbreviation.
In the case of $n=1$, it is easily shown
\ba
\label{eq:1tripmag-1}
\ket{S_{k}^{+}} &=&\sum_i e^{ikx_i}S^{+}_i\ket{A}\nn
&=&\sum_i e^{ikx_i}\ket{{\cal T}^{1}_{i-1}} - \sum_i e^{ikx_i}\ket{{\cal T}^{1}_i}\nn
&=&(e^{ik}-1)\ket{{\cal T}_{k}^{1}}.
\ea
Assuming that the equivalence holds between the $n$-magnon and $n$-triplon states with momenta $k_2$ through $k_{p+1}$, {\it i.e.} $\ket{S_{k_2}^{+}\cdots S_{k_{p+1}}^{+}}= \left[ \Pi_{j=2}^{p+1} P_j  \right] \ket{{\cal T}_{k_2}^{1}\cdots {\cal T}_{k_{p+1}}^{1}}$, the $n=p+1$ magnon state is defined as
\ba\label{eq:EMT} && \ket{S_{k_1}^{+}S_{k_2}^{+}\cdots S_{k_{p+1}}^{+}}  = \sum_{i_1} e^{ik_1 x_{i_1}} S^+_{i_1} \ket{S_{k_2}^{+}\cdots S_{k_{p+1}}^{+}}\nn
&=&[ \Pi_{j=2}^{p+1}P_j ] \sum_{i_1,\cdots,i_{p+1}} e^{\sum_{j=1}^{p+1} i k_j x_{i_j}}{\cal S}_{i_1}^+\ket{{\cal T}^{1}_{i_2}\cdots{\cal T}^{1}_{i_{p+1}}} .
\ea
The operator $S_{i_1}^+$ acting on $\ket{{\cal T}^{1}_{i_2}\cdots{\cal T}^{1}_{i_{p+1}}}$ becomes
\ba
S_{i_1}^+\ket{{\cal T}^{1}_{i_2}\cdots{\cal T}^{1}_{i_{p+1}}}&=&\ket{{\cal T}_{i_{1}-1}^{1}{\cal T}^{1}_{i_2}\cdots{\cal T}^{1}_{i_{p+1}}}-\ket{{\cal T}_{i_1}^{1}{\cal T}^{1}_{i_2}\cdots{\cal T}^{1}_{i_{p+1}}}\nn\label{eq:ST1}
\ea
when none of the indices from $i_2$ through $i_{p+1}$ coincides with $i_1$ or $i_1-1$.
Since arbitrary permutation of the indices in the $n$-triplon state $\ket{{\cal T}^{1}_{i_2}\cdots{\cal T}^{1}_{i_{p+1}}}$ gives the same state, we only need to consider the following three situations: $i_2 = i_1 -1$, $i_2 = i_1$, and $i_2 = i_1 -1, i_3 = i_1$. In each case, we have
\ba
S_{i_1}^+\ket{{\cal T}_{i_1-1}^{1}\cdots}&=&-\ket{{\cal T}_{i_1-1}^{1}{\cal T}_{i_1}^{1}\cdots}\nn
S_{i_1}^+\ket{{\cal T}_{i_1}^{1}\cdots}&=&\ket{{\cal T}_{i_1-1}^{1}{\cal T}_{i_1}^{1}\cdots}\nn
S_{i_1}^+\ket{{\cal T}_{i_1-1}^{1}{\cal T}_{i_1}^{1}\cdots}&=&0.\label{eq:ST2}
\ea
Using Eqs. (\ref{eq:ST1}) and (\ref{eq:ST2}), Eq.(\ref{eq:EMT}) becomes
\begin{widetext}
\ba
({\rm B3})&=& [ \Pi_{j=2}^{p+1} P_j ] \Bigl(\sum'_{\{i\}} e^{\sum_{j=1}^{p+1} i k_j x_{i_j}}\ket{{\cal T}^{1}_{i_1-1}{\cal T}^{1}_{i_2}\cdots{\cal T}^{1}_{i_{p+1}}}-\sum''_{\{i\}} e^{\sum_{j=1}^{p+1} i k_j x_{i_j}}\ket{{\cal T}^{1}_{i_1}{\cal T}^{1}_{i_2}\cdots{\cal T}^{1}_{i_{p+1}}}\nn
&&+\sum_{i_1,\cdots,i_{p+1}}\sum_{m=2}^{p+1}e^{\sum_{j=1}^{p+1} i k_j x_{i_j}} (\delta_{i_m,i_1}\ket{{\cal T}^{1}_{i_1-1}{\cal T}^{1}_{i_2}\cdots{\cal T}^{1}_{i_{p+1}}}-\delta_{i_m,i_1-1}\ket{{\cal T}^{1}_{i_1}{\cal T}^{1}_{i_2}\cdots{\cal T}^{1}_{i_{p+1}}})\Bigl).
\ea

In the first sum $\sum'_{\{i\}}$, indices $i_2, \cdots, i_{p+1}$ run through all sites except $i_1$.
In the second sum $\sum''_{\{i\}}$, indices $i_2, \cdots, i_{p+1}$ overlapping with $i_1 -1$ are excluded. Remarkably the third and the fourth term can be merged with the first and the second term and become triplon wave states, respectively. Therefore,
\ba
({\rm B6} ) &=&[ \Pi_{j=2}^{p+1}P_j ] \sum_{i_1,\cdots,i_{p+1}} e^{\sum_{j=1}^{p+1} i k_j x_{i_j}}(\ket{{\cal T}^{1}_{i_1-1}{\cal T}^{1}_{i_2}\cdots{\cal T}^{1}_{i_{p+1}}} -\ket{{\cal T}^{1}_{i_1}{\cal T}^{1}_{i_2}\cdots{\cal T}^{1}_{i_{p+1}}})\nn
&=& [ \Pi_{j=1}^{p+1} P_j ] \ket{{\cal T}_{k_1}^{1}{\cal T}_{k_2}^{1}\cdots {\cal T}_{k_{p+1}}^{1}}.
\ea
\end{widetext}
In the same way one can prove the equivalence of $\ket{S_{k_1}^{-}S_{k_2}^{-}\cdots S_{k_n}^{-}}$ and $\ket{{\cal T}_{k_1}^{-}{\cal T}_{k_2}^{-}\cdots {\cal T}_{k_n}^{-}}$.

\section{Denominator of the two-triplon SMA}
\label{appendix:denominator}
To begin, let us consider the simplest case with $\alpha_1 = \alpha_2 =1$. We are trying to compute
\ba
&& \langle {\cal T}_{k_{1}}^{1}{\cal T}_{k_{2}}^{1}|{\cal T}_{k_{1}}^{1}{\cal T}_{k_{2}}^{1} \rangle
\nn
\label{eq:tripdensma}
&& = \sum_{i,j,k,m} e^{ik_1(x_i-x_k) + ik_2(x_j-x_m)} \langle {\cal T}_k^1 {\cal T}_m^1 | {\cal T}_i^1 {\cal T}_j^1 \rangle .
\ea
The main idea we use is to break the above unrestricted sum over indices into sums over indices satisfying inequality constraints. The MPS overlap method as detailed in Appendix \ref{appendix:TM} then allows us to compute
$\langle {\cal T}_k^1 {\cal T}_m^1 | {\cal T}_i^1 {\cal T}_j^1 \rangle$ for a particular ordered quadruple $i,j,k,m$ and the Fourier sum can then be exactly computed. The following symmetries of the overlap can be exploited to
reduce the number of required computations.

Since in general we have $| {\cal T}_i^{1} {\cal T}_j^{1} \rangle = | {\cal T}_j^{1} {\cal T}_i^{1} \rangle$,
the overlap   must satisfy
\begin{multline}  \langle {\cal T}_k^{1} {\cal T}_m^{1} | {\cal T}_i^{1} {\cal T}_j^{1} \rangle = \langle {\cal T}_m^{1} {\cal T}_k^{1} | {\cal T}_i^{1} {\cal T}_j^{1} \rangle \\
= \langle {\cal T}_k^{1} {\cal T}_m^{1} | {\cal T}_j^{1} {\cal T}_i^{1} \rangle  = \langle {\cal T}_m^{1} {\cal T}_k^{1} | {\cal T}_j^{1} {\cal T}_i^{1} \rangle .
\end{multline}
As a result, we can assume $k < m$ and $i<j$ without loss of generality.  Furthermore, there is another identity of the overlap,
\ba \langle {\cal T}_k^{1} {\cal T}_m^{1} | {\cal T}_i^{1} {\cal T}_j^{1} \rangle = \langle {\cal T}_i^{1} {\cal T}_j^{1} | {\cal T}_k^{1} {\cal T}_m^{1} \rangle , \ea
which allows us to restrict ourselves to $k \le i$ without loss of generality. There are then only three cases to
consider: (i) both $(i,j)$ lie inside the $[k, m]$ interval, (ii) both $(i,j)$ lie outside the $[k,m]$ interval, and
(iii) only $i$ lies inside the $[k,m]$ interval. In the actual calculation below, we reorganize this to four cases for computational facility.

The overlaps computed by the transfer matrix method as decribed in  the Appendix \ref{appendix:TM},  for the different cases relevant to the denominator of the two-triplon SMA for the
$\alpha_1 =\alpha_2 =1$ case are given by
\newline

\noindent
{\bf Case 1: $k \leq i < m < j$}: In this case only $i$ lies inside the $[k,m)$ interval. We have in the thermodynamic limit,
\ba
\label{eq:full-ovl}
\langle {\cal T}_k^{1} {\cal T}_m^{1} | {\cal T}_i^{1} {\cal T}_j^{1}\rangle&=& \frac{1}{4} 3^N \left(  (-\frac{1}{3})^{i+j-k-m} - (-\frac{1}{3})^{j-k} \right).\nn
\ea
{\bf Case 2: $k < i <j\leq m$}: This is the situation where both $i$ and $j$ lie inside the interval $(k,m]$. We obtain
\ba
\langle {\cal T}_k^{1} {\cal T}_m^{1} | {\cal T}_i^{1} {\cal T}_j^{1} \rangle &=& \frac{1}{4} 3^N  \left( (-\frac{1}{3})^{i-j-k+m}  -(-\frac{1}{3})^{-k+m} \right).\nn
\ea
{\bf Case 3: $k<m\leq i <j$}: This is case where both $(i,j)$ lie outside the $[k,m)$ range. The overlap vanishes in the thermodynamic limit.
%
%
\\
{\bf Case 4: $k=i < m =j$}: In this case, there are only two independent indices. We obtain
\ba
\langle {\cal T}_k^{1} {\cal T}_m^{1} | {\cal T}_k^{1} {\cal T}_m^{1} \rangle &=& \frac{1}{4} 3^N \left[ 1- (-\frac{1}{3})^{-k+m}  \right].
\ea
\newline

The summation in Eq. (\ref{eq:tripdensma}) can be broken up into 24 summations over inequalities like $k<i<m<j$, $k<i<j<m$ and so on. In addition, one has to consider situations when some
of these indices become equal. The overlap formulas mentinoned above covers  all the distinct type of overlap calculations by the transfer matrix method that we need to do. The overlap required
for any other case, not belonging to the four cases above, can be obtained from one of the formulas above by interchanging some indices due to the symmetries we mentioned before.
We refer to such inequalities as subcases within a case. Because of the presence of the phase in Eq. (\ref{eq:tripdensma}), the results of the sums for subcases
within a case are different.
There are eight sub-cases for cases 1, 2, and 3, and  four sub-cases for case 4  considering
different permutations of the indices $i,j,k,m$. Let us define the sums over subcases as,
\ba
S^{1}_{abcd} &=& \sum_{a \leq b <c <d}   e^{ik_1 (x_i-x_k) + ik_2(x_j-x_m)}  \langle {\cal T}_k^{1} {\cal T}_m^{1} | {\cal T}_i^{1} {\cal T}_j^{1} \rangle \nn
S^{2}_{abcd} &=& \sum_{a < b <c \leq d}   e^{ik_1 (x_i-x_k) + ik_2(x_j-x_m)}  \langle {\cal T}_k^{1} {\cal T}_m^{1} | {\cal T}_i^{1} {\cal T}_j^{1} \rangle \nn
S^{3}_{abcd} &=& \sum_{a < b \leq c <d}   e^{ik_1 (x_i-x_k) + ik_2(x_j-x_m)}  \langle {\cal T}_k^{1} {\cal T}_m^{1} | {\cal T}_i^{1} {\cal T}_j^{1} \rangle \nn
S^{4}_{abcd} &=& \sum_{a = b <c = d}   e^{ik_1 (x_i-x_k) + ik_2(x_j-x_m)}  \langle {\cal T}_k^{1} {\cal T}_m^{1} | {\cal T}_i^{1} {\cal T}_j^{1} \rangle. \nn
\ea
On using the overlaps computed above for case 1 we get, to the leading order,
\begin{eqnarray}
S^1_{kimj}&=&-\frac{3}{8} { e^{i k_2}  \over (  3+e^{i k_1} ) ( 3+e^{i k_2}  ) }  3^N N^2\nn
S^1_{ikjm} &=& -\frac{3}{8} \frac{ e^{i k_1}}{ \left(1+3 e^{i k_1}\right) \left(1+3 e^{i k_2}\right)} 3^N N^2 \nn
S^1_{mikj} &=& S^1_{kjmi} = -\frac{3}{8}  \frac{ e^{i k_1}}{ \left(3+e^{i k_1}\right)^2} \delta_{k_1,k_2} 3^N N^2 \nn
S^1_{jkim} &=& S^1_{imjk} = -\frac{3}{8} \frac{ e^{i k_1}}{ \left(1+3 e^{i k_1}\right)^2 }  \delta_{k_1,k_2} 3^N N^2.
\end{eqnarray}
$S^1_{mjki}$ and $S^1_{jmik}$ can be obtained respectively from $S^1_{kimj}$ and $S^1_{ikjm}$ after interchanging $k_1$ and $k_2$. For case 2  we have,
\begin{eqnarray}
S^2_{kijm} &=& -\frac{3}{8} \frac{ e^{i \left(k_1+k_2\right)}}{ \left(3+e^{i k_1}\right) \left(1+3 e^{i k_2}\right)} 3^N N^2 \nn
S^2_{ikmj} &=& -\frac{3}{8} \frac{1}{ \left(1+3 e^{i k_1}\right) \left(3+e^{i k_2}\right)} 3^N N^2 \nn
S^2_{mijk} &=& S^2_{kjim} = -\frac{3}{8}  \frac{ e^{ 2 i k_1 }}{\left(3+e^{i k_1}\right) \left(1+3 e^{i k_1}\right)}  \delta_{k_1,k_2} 3^N N^2 \nn
S^2_{imkj} &=& S^2_{jkmi} =  -\frac{3}{8} \frac{1}{\left(1+3 e^{i k_1}\right) \left(3+e^{i k_1}\right)} \delta_{k_1,k_2} 3^N N^2.\nn
\end{eqnarray}
$S^2_{mjik}$ and $ S^2_{jmki}$ can be obtained respectively from $S^2_{kijm}$ and $S^2_{ikmj}$ after interchanging $k_1$ and $k_2$.
For case 4, we obtain
\ba
S^4_{kimj}=S^4_{mjki}&=&\frac{1}{8} 3^N N^2 \nn
S^4_{kjmi}=S^4_{mikj}&=&\frac{1}{8} \delta_{k_1,k_2} 3^N N^2.
\ea
All the other subcases in case 1, case 2, and case 4 contribute lower order terms and also all subcases in case 3.
On adding all these contributions and assuming $N$ is large, we get
\begin{equation}
\langle {\cal T}_{k_{1}}^{1}{\cal T}_{k_{2}}^{1}|{\cal T}_{k_{1}}^{1}{\cal T}_{k_{2}}^{1} \rangle   \approx
\frac{ 4(1 + \delta_{k_1,k_2})}{(5+3\cos{k_1})(5+3\cos{k_2})} 3^N N^2.
\end{equation}

In the case of $\alpha \neq \beta$, there is only one symmetry we can use,
\ba \langle {\cal T}_k^{\alpha} {\cal T}_m^{\beta} | {\cal T}_i^{\alpha} {\cal T}_j^{\beta} \rangle = \langle {\cal T}_i^{\alpha} {\cal T}_j^{\beta} |
{\cal T}_k^{\alpha} {\cal T}_m^{\beta} \rangle
\ea
Therefore, we can say $k\leq i$ without breaking generality. The rest of the steps for other spin indices are similar to the  $\alpha = \beta=1$ case, although we have
to compute more overlaps compared to the latter due to reduced symmetry.
\section{Numerator of the  two-triplon SMA}
\label{appendix:numerator}
The numerator of the two-triplon SMA is given by
\ba
\label{eq:trip-num1}
&&\langle {\cal T}_{k_1}^{1} {\cal T}_{k_2}^{1} |H_A| {\cal T}_{k_1}^{1} {\cal T}_{k_2}^{1}\rangle\nn
&&=\sum_{i,j,k,m}e^{ik_1(x_i-x_k) + ik_2(x_j-x_m)} \langle {\cal T}_k^{1} {\cal T}_m^{1}| H_A|{\cal T}^{1}_i {\cal T}^{1}_j \rangle.\nn
\ea
On using the fact that $H_i$ annihilates any triplon state  with a singlet operator at site $i$ in the Schwinger boson representation, we obtain
\begin{widetext}
\ba
\langle {\cal T}^{1}_k {\cal T}^{1}_m| H_A|{\cal T}^{1}_i {\cal T}^{1}_j \rangle &=&  \langle {\cal T}^{1}_k {\cal T}^{1}_m| (H_i + H_j)|{\cal T}^{1}_i {\cal T}^{1}_j \rangle\nn
&=& (\delta_{k,i} +\delta_{m,i})\langle {\cal T}^{1}_k {\cal T}^{1}_m| H_i |{\cal T}^{1}_i {\cal T}^{1}_j \rangle +(\delta_{m,j}+\delta_{k,j}) \langle {\cal T}^{1}_k {\cal T}^{1}_m| H_j |{\cal T}^{1}_i {\cal T}^{1}_j \rangle.
\ea
Putting this expression for $\langle {\cal T}^{1}_k {\cal T}^{1}_m| H_A|{\cal T}^{1}_i {\cal T}^{1}_j \rangle$  back to Eq.~(\ref{eq:trip-num1}) we have
\ba
\label{eq:twotrip-num}
\langle {\cal T}_{k_1}^{1} {\cal T}_{k_2}^{1} |H_A| {\cal T}_{k_1}^{1} {\cal T}_{k_2}^{1}\rangle&=& \sum_{i,j,k ;j,k \ne i}(e^{ik_2(x_j-x_k)} +  e^{ik_1(x_i-x_k) + ik_2(x_j-x_i)})\langle {\cal T}^{1}_i{\cal T}^{1}_k| H_i |{\cal T}^{1}_i {\cal T}^{1}_j \rangle\nn
&+& \sum_{i,j,k ;j,k \ne i}(e^{ik_1(x_j-x_k)} +  e^{ik_2(x_i-x_k) + ik_1(x_j-x_i)})  \langle {\cal T}^{1}_i{\cal T}^{1}_k| H_i |{\cal T}^{1}_i {\cal T}^{1}_j \rangle.\nn
\ea
\end{widetext}
The numerator can now be calculated by performing the Fourier sum in the same way as the denominator after computing the overlaps $\langle {\cal T}^{1}_i{\cal T}^{1}_m| H_i |{\cal T}^{1}_i {\cal T}^{1}_j \rangle$. We also require the following  results of the action of  the AKLT Hamiltonian on the two triplon states computed using standard Schwinger-Boson techniques,
\begin{eqnarray}
\label{eq:aklt-H-trip}
H_i |{\cal T}^1_i {\cal T}^1_j \rangle &=& \frac{1}{24} (4|{\cal T}^1_{i-1} {\cal T}^0_{i+1} {\cal T}^1_{j}\rangle + 8|{\cal T}^1_{i-1}{\cal T}^1_{j} \rangle + 8|{\cal T}^1_{i+1} {\cal T}^1_{j} \rangle\nn
&-4&|{\cal T}^0_{i-1} {\cal T}^1_{i+1} {\cal T}^1_{j} \rangle+ 24 |{\cal T}^1_{i} {\cal T}^1_{j} \rangle ) ,\: j \ne i-1,i+1 \nn
H_i |{\cal T}^1_i {\cal T}^1_{i+1} \rangle &=& \frac{1}{2} |{\cal T}^1_{i-1} {\cal T}^1_{i+1} \rangle +  |{\cal T}^1_i {\cal T}^1_{i+1} \rangle \nonumber  \\
H_i |{\cal T}^1_i {\cal T}^1_{i-1} \rangle &=& \frac{1}{2} |{\cal T}^1_{i-1} {\cal T}^1_{i+1} \rangle +  |{\cal T}^1_i {\cal T}^1_{i-1} \rangle.
\end{eqnarray}

An important point to note is that when computing overlap between two states with even and odd number of triplons respectively or vice versa, an extra negative sign
has to be considered in addition to the transfer matrix result (see Eq. (\ref{eq:rep-rule})). Moreover the 
term with the second summation sign  on the right hand side of Eq.~\ref{eq:twotrip-num} can be derived from the  term with the first summation sign after exchanging $k_1$ and $k_2$.

As before with the denominator, the strategy is to break the summation $\sum_{i,j,k ;j,k \ne i}$ into different sums $\sum_{i<j<k} + \sum_{j<k<i}$ and so on, six terms in total and an equality case in addition. This gives rise to the cases decribed below.

Let us define $\phi(i,j,k)$ as
\ba
\phi(i,j,k)&=&e^{ik_1(x_i-x_k) + ik_2(x_j-x_i)}+e^{ik_2(x_j-x_k)}
\ea
and $\psi_n$ as $\langle {\cal T}^1_k{\cal T}^1_i| H_i|{\cal T}^1_i {\cal T}^1_j \rangle$ for case $n$. 
For each case, we put equations of $\psi_n$ below for  \{i,j,k\} belonging to disjoint sets. Since the matrix element, $ \langle {\cal T}^1_k{\cal T}^1_i| H_i|{\cal T}^1_i {\cal T}^1_j \rangle $ is real we have,
\begin{equation}
\langle {\cal T}^1_k{\cal T}^1_i| H_i|{\cal T}^1_i {\cal T}^1_j \rangle =  \langle {\cal T}^1_k{\cal T}^1_i| H_i|{\cal T}^1_i {\cal T}^1_j \rangle^*=
\langle {\cal T}^1_i{\cal T}^1_j| H_i|{\cal T}^1_i {\cal T}^1_k \rangle.
\end{equation}
Hence for a subcase such as $j<i<k$, the relevant overlaps can be obtained from the overlaps computed below for the case $k<i<j$  by interchanging $j$ and $k$. The rest of the calculation will proceed in an exactly similar manner.
\newline

\noindent
{\bf Case 1: $k<i<j$}\\
As the tranfer matrix  calculation $\langle {\cal T}^1_k{\cal T}^1_i| H_i|{\cal T}^1_i {\cal T}^1_j \rangle$ becomes different for $k=i-1$ or $j=i+1$ we need to consider
these terms separately. We have
\begin{widetext}
\ba
\sum_{k<i<j}\phi(i,j,k) \psi_1 (i,j,k) &=& \sum_{k<i<j ; k \ne i-1, j \ne i+1}\phi(i,j,k)\psi_1(i,j,k)+ \sum_{i<j-1} \phi(i,j,i-1) \psi_1(i,j,i-1)\nn
&&+\sum_{k<i-1} \phi(i,i+1,k) \psi_1(i,i+1,k) +   \sum_i e^{i(k_1+k_2)}\psi_1(i,i+1,i-1).
\ea
\end{widetext}
We have on calculating the overlaps using   Eq.(\ref{eq:aklt-H-trip}) in the large-$N$ limit,
\ba
\psi_1(i,j,k) &=& \frac{1}{54} (-1)^{j-k+1} 3^{-j+k+N+3}\nn
\psi_1(i,j,i-1) &=& \frac{1}{6} (-1)^{j-i} 3^{i-j+N}\nn
\psi_1(i,i+1,k) &=& \frac{1}{6} (-1)^{i-k} 3^{-i+k+N}\nn
\psi_1(i,i+1,i-1) &=& -\frac{1}{18}3^N.
\ea
{\bf Case 2: $i<j<k$}\\
We have,
\ba
\sum_{i<j<k}  \phi(i,j,k) \psi_2 (i,j,k) &=& \sum_{i<j<k ;  j \ne i+1} \phi(i,j,k)   \psi_2(i,j,k)\nn
&+&\sum_{i<k-1} \phi(i,i+1,k) \psi_2(i,i+1,k)\nn
\ea
The necessary overlaps  calculated  by using Eq. (\ref{eq:aklt-H-trip}) in the large-N limit are given by,
\ba
\psi_2(i,j,k) &=& \frac{1}{54} 3^N \left(-15 (-\frac{1}{3})^{k-i} + 10 (-\frac{1}{3})^{k-j}  \right)\nn
\psi_2(i,i+1,k) &=& \frac{1}{6} 3^N\left( -5 (-\frac{1}{3})^{k-i}\right).
\ea
{\bf Case 3:  $k<j<i$}\\
We have,
\ba
\sum_{k<j<i }  \phi(i,j,k) \psi_3(i,j,k)&=& \sum_{k<j<i ;  j \ne i-1} \phi(i,j,k)   \psi_3(i,j,k)\nn
&+&\sum_{k<i-1 } \phi(i,i-1,k) \psi_3(i,i-1,k).\nn
\ea
The necessary overlaps  calculated using Eq. (\ref{eq:aklt-H-trip}) in the large-N limit are given by,
\ba
\psi_3(i,j,k) &=&  \frac{1}{54} 3^N \left(  -15 (-\frac{1}{3})^{i-k} + 10 (-\frac{1}{3})^{j-k} \right)\nn
\psi_3(i,i-1,k) &=& \frac{1}{6} 3^N\left(   -5 (-\frac{1}{3})^{i-k} \right).
\ea
{\bf Case 4: $j=k$}\\
We have,
\ba
&&\sum_{i,j \ne i }  \phi(i,j,j)  \psi_4(i,j,k)  = \sum_{i,j \ne i,i-1,i+1} \phi(i,j,j)   \psi_4(i,j,j)+ \nn
&&\sum_{i} \left(e^{i(k_1-k_2)}  \psi_4(i+1,i,i)+e^{-i(k_1-k_2)} \psi_4(i-1,i,i) \right).\nn
\ea
The necessary overlaps  calculated using Eq. (\ref{eq:aklt-H-trip}) in the large-N limit are given by,
\begin{equation}
\psi_4(i,j,j)= \frac{1}{54} 3^N \left(  -15 (-\frac{1}{3})^{j-i}  +  10  \right).
\end{equation}

It also turns out after calculation that $\psi_4(i,j,j)$ for $j<i$ can be obtained from above by interchanging $i,j$.


Let us now describe our results. It turns out that only $\sum_{i,j,k ;j,k \ne i}e^{ik_2(x_j-x_k)}  \langle {\cal T}^{1}_i{\cal T}^{1}_k| H_i |{\cal T}^{1}_i {\cal T}^{1}_j \rangle$ for case 1, 2, 3 produce  ${\mathcal{O}}(3^N N^2)$ terms. This is perhaps not surprising since already in the denominator calculation we saw that the contributing sums are those which have substantial overlap between the triplonic bonds, i.e, between $i$, $k$ and $i$, $j$ in this case. The results of the remaining sums at points where the denominator of the sums are non-zero are found to be,
\begin{eqnarray}
I_{ijk} &=& I_{jki} = -\frac{5}{54 \left(1+3 e^{i k_2}\right)}3^N N^2  \nonumber \\
I_{ikj}  &=& I_{kji} = -\frac{5 e^{i k_2}}{54 \left(3+e^{i k_2}\right)}3^N N^2  \nonumber \\
I_{eq} &=& \frac{5}{27}3^N N^2.
\end{eqnarray}
Combining and considering  terms coming from second term of Eq. (\ref{eq:twotrip-num}) we have the numerator(for $k_1 \ne k_2$) equal to,
$\left(\frac{20}{81 \cos \left(k_2\right)+135} + \frac{20}{81 \cos \left(k_1\right)+135}\right)3^N N^2$.

Earlier we found that the denominator of the two-triplon SMA  have discontinuities at  $k_1=k_2$. The numerator also becomes exactly double
of its value at $k_1=k_2$ compared to other points in the vicinity, as the two phases within the first(second) summation in the right hand side of 
Eq.(\ref{eq:twotrip-num}) become equal to each other.
The calculation for other values of $\alpha, \beta$ proceeds in a similar way, but as before with the denominator we have to compute more overlaps due to reduced symmetry.
For example for $\alpha=1, \beta=0$ we need to compute,
\begin{widetext}
\ba
\label{eq1:twotrip-num}
\langle {\cal T}_{k_1}^{1} {\cal T}_{k_2}^{0} |H_A| {\cal T}_{k_1}^{1} {\cal T}_{k_2}^{0}\rangle 
&=& \sum_{i,j,m ;j,m \ne i}\Bigl(e^{ik_2(x_j-x_m)}  \langle {\cal T}^{1}_i{\cal T}^{0}_m| H_i |{\cal T}^{1}_i {\cal T}^{0}_j \rangle+e^{ik_1(x_j-x_m)} \langle {\cal T}^{1}_m{\cal T}^{0}_i| H_i|{\cal T}^{1}_j {\cal T}^{0}_i \rangle\nn
&&+ e^{ik_1(x_i-x_m)+ik_2(x_j-x_i)}\langle {\cal T}^{1}_m{\cal T}^{0}_i| H_i|{\cal T}^{1}_i {\cal T}^{0}_j \rangle+ 
e^{ik_2(x_i-x_m) + ik_1(x_j-x_i)}\langle {\cal T}^{1}_i{\cal T}^{0}_m| H_i|{\cal T}^{0}_i {\cal T}^{1}_j \rangle\Bigl).
\ea
\end{widetext}
As in other cases, only the first two terms of the last line of the above equation contribute to the result metioned in the main text.
However, unlike for $\alpha= \beta$ cases there is no discontinuity at $k_1=k_2$,
as the first (second) term of 
Eq. (\ref{eq1:twotrip-num}) {\emph{does not}} become equal to  the
third (fourth) term, even though the phases both become equal to $e^{ik_2(x_j-x_m)}$, because
$ \langle {\cal T}^{1}_i{\cal T}^{0}_m| H_i |{\cal T}^{1}_i {\cal T}^{0}_j \rangle \ne \langle {\cal T}^{1}_m{\cal T}^{0}_i| H_i|{\cal T}^{1}_i {\cal T}^{0}_j \rangle $.

\bibliography{triplons-aklt}

\end{document}